\documentclass[aps,twocolumn,showpacs]{revtex4}
\usepackage{amsgen}
\usepackage{amsfonts}
\usepackage{amssymb}
\usepackage{amsbsy}
\usepackage{epsfig}
\usepackage{bm}
\usepackage{color}
\usepackage{ulem}
\usepackage{dcolumn}
\usepackage{dsfont}

\begin{document}

%
%
%
%
\def\ket#1{ $ \left\vert  #1   \right\rangle $ }
\def\ketm#1{  \left\vert  #1   \right\rangle   }
\def\bra#1{ $ \left\langle  #1   \right\vert $ }
\def\bram#1{  \left\langle  #1   \right\vert   }
\def\spr#1#2{ $ \left\langle #1 \left\vert \right. #2 \right\rangle $ }
\def\sprm#1#2{  \left\langle #1 \left\vert \right. #2 \right\rangle   }
\def\me#1#2#3{ $ \left\langle #1 \left\vert  #2 \right\vert #3 \right\rangle $ }
\def\mem#1#2#3{  \left\langle #1 \left\vert  #2 \right\vert #3 \right\rangle   }
\def\redme#1#2#3{ $ \left\langle #1 \left\Vert
                  #2 \right\Vert #3 \right\rangle $ }
\def\redmem#1#2#3{  \left\langle #1 \left\Vert
                  #2 \right\Vert #3 \right\rangle   }
\def\threej#1#2#3#4#5#6{ $ \left( \matrix{ #1 & #2 & #3  \cr
                                           #4 & #5 & #6  } \right) $ }
\def\threejm#1#2#3#4#5#6{  \left( \matrix{ #1 & #2 & #3  \cr
                                           #4 & #5 & #6  } \right)   }
\def\sixj#1#2#3#4#5#6{ $ \left\{ \matrix{ #1 & #2 & #3  \cr
                                          #4 & #5 & #6  } \right\} $ }
\def\sixjm#1#2#3#4#5#6{  \left\{ \matrix{ #1 & #2 & #3  \cr
                                          #4 & #5 & #6  } \right\} }

\def\ninejm#1#2#3#4#5#6#7#8#9{  \left\{ \matrix{ #1 & #2 & #3  \cr
                                                 #4 & #5 & #6  \cr
                         #7 & #8 & #9  } \right\}   }
%
%
%
%

\newcommand{\pr}{^{\prime}}
\newcommand{\bfx}{\overrightarrow{x}}
\newcommand{\bfp}{{\bf p}}
\newcommand{\la}{\langle}
\newcommand{\ra}{\rangle}
\newcommand{\rp}{{p}}
\newcommand{\rpp}{{{p}^{\prime}}}
\newcommand{\rx}{{x}}
\newcommand{\vare}{\varepsilon}
\newcommand{\eps}{\varepsilon}
\newcommand{\be}{\begin{eqnarray}}
\newcommand{\ee}{\end{eqnarray}}
\newcommand{\ba}{\begin{array}}
\newcommand{\ea}{\end{array}}
\newcommand{\proc}{{alpha decay }}
\newcommand{\balpha}{{\mbox{\boldmath$\alpha$}}}
\newcommand{\bsigma}{{\mbox{\boldmath$\sigma$}}}
\newcommand{\bnabla}{{\mbox{\boldmath$\nabla$}}}
\newcommand{\bfr}{{\bf r}}
\newcommand{\bfb}{{\bf b}}
\newcommand{\bfe}{{\bf e}}
\newcommand{\bfk}{{\bf k}}
\newcommand{\bfK}{{\bf K}}
\newcommand{\bfR}{{\bf R}}
\newcommand{\bfA}{{\bf A}}
\newcommand{\bbeta}{{\mbox{\boldmath$\beta$}}}
\newcommand{\bomega}{{\mbox{\boldmath$\omega$}}}
\newcolumntype{.}{D{.}{.}{-1}}
\newcolumntype{d}[1]{D{.}{.}{#1}}


%
%
\title{Solution of two--center time--dependent Dirac equation in spherical coordinates: \\[0.2cm]
Application of the multipole expansion of the electron--nuclei interaction}

%
%
\author{S.~R.~McConnell\footnote{Corresponding author. Email:
    smcconne@physi.uni-heidelberg.de}$^{1,2}$, A.~N.~Artemyev$^{1,2}$,
  M.~Mai$^{1,3}$ and  A.~Surzhykov$^{1,2}$}
\affiliation{$^{1}$
Physikalisches Institut, Universit\"{a}t Heidelberg, Im Neuenheimerfeld 226, D--69120 Heidelberg, Germany\\
$^{2}$GSI Helmholtzzentrum f\"ur Schwerionenforschung GmbH, Planckstr. 1, D--64291 Darmstadt, Germany\\
$^{3}$Department of Physics, Yale University, 217 Prospect Street, New Haven, Connecticut 06511--8499, USA
}
\date{\today}

%
%
%
%
\begin{abstract}
A non--perturbative approach to the solution of the time--dependent, two--center Dirac equation is presented with a special emphasis on the proper treatment of the potential of the nuclei. In order to account for the full multipole expansion of this potential, we express eigenfunctions of the two--center Hamiltonian in terms of well--known solutions of the ``monopole'' problem that employs solely the spherically--symmetric part of the interaction. When combined with the coupled--channel method, such a wavefunction--expansion technique allows for an accurate description of the electron dynamics in the field of moving ions for a wide range of internuclear distances. To illustrate the applicability of the proposed approach, the probabilities of the $K$-- as well as $L$-- shell ionization of hydrogen--like ions in the course of nuclear $\alpha$-decay and slow ion--ion collisions have been calculated.
\end{abstract}

\pacs{31.30.Jv, 34.80.Dp, 34.50.Fa}
\maketitle

%
%
%
%
\section{Introduction}

Recent developments in accelerator and storage ring technologies have made it possible to perform a new generation of experiments on collisions between heavy, highly--charged ions. Of special interest in these studies are the \textit{low--energy} collisions leading to the formation of short--lived quasi--molecular systems in which electrons move in the Coulomb field of two (or more) nuclei. Analysis of the excitation, ionization, charge--transfer, and pair--production processes in such a low--energy domain may reveal important information about the properties and behaviour of few--electron systems and even of the quantum vacuum in the presence of extremely strong electromagnetic fields. To achieve and exploit the strong--field regime, a broad research program is planned to be undertaken at the future Facility for Antiproton and Ion Research (FAIR) in Darmstadt, at which ions up to bare uranium will be produced and decelerated to required energies \cite{FAIR01,GuS09}.

\medskip

In order to better understand the basic atomic processes accompanying slow ion collisions, the experimental findings have to be supplemented by a detailed theoretical analysis. In the simplest case of the collision between bare and hydrogen--like heavy ions, such an analysis can be traced back to the single--electron two--center Dirac problem. For small relative velocities and comparable charges of the nuclei, $Z_1 \simeq Z_2$, the \textit{non--perturbative} treatment of such a problem is usually required and can be performed by using various coupled--channel techniques. Along this line, the time--dependent electron wave packet is expanded in terms of eigensolutions of the stationary Dirac equation, which describes the two--center system at a fixed internuclear distance $R$. The performance of the coupled--channel methods depends, therefore, on the efficiency of the spectrum generation of the time--independent Hamiltonian at each required $R$.

\medskip

An accurate solution of the static two--center problem is in general a rather sophisticated task which can benefit from a proper choice of coordinate system. During the last two decades in particular, a number of theoretical methods have been developed which make use of Cassini \cite{ScW83,WiS83,ArS10} and prolate spheroidal \cite{KuK01,FiG12} coordinate systems. Even though these (non--spherical) coordinates are very practical for the computation of quasi--molecular spectra at arbitrary internuclear distance $R$, their employment may be hampered by the lack of established numerical techniques for the evaluation of two--center matrix elements. Consequently, retention of standard \textit{spherical coordinates} for the treatment of ion--ion (or ion--atom) collisions still attracts much current attention. The use of these, essentially one--center, coordinates for the description of the two--center problem also requires the development of various approximate methods. Within the linear combination of atomic orbitals (LCAO) approaches \cite{SeK86,MoG93,GaG03,TuK10,TuK12}, for example, quasi--molecular wavefunctions are constructed from sets of atomic orbitals, centered on each nucleus. Yet another and very promising method relies on the \textit{direct} solution of the two--center Dirac problem. Such a solution is rather straightforward and well--elaborated if the electron--nuclei potential is approximated by its spherically symmetrical part \cite{BeS76,ReM81,AcH07,AcH08,DeM12}. This so--called monopole approximation is successfully used for the description of strong--field phenomena in close--ion collisions, but performs poorly when the Coulomb centers are far from each other. The extension of the multipole theory towards accounting for higher terms in the decomposition of the two--center potential is crucial, therefore, for the proper treatment of heavy--ion collisions in spherical coordinates.

\medskip

A number of efforts have been focused in the past on a straightforward solution of the (radial) Dirac equation for the complete two--center potential \cite{Gre00,MaH11}. In these studies, the components of the quasi--molecular wavefunctions were found upon integration of an infinite system of coupled differential equations which account for all terms of multipole expansion. An alternative and computationally very efficient approach to the two--center problem in spherical coordinates is proposed in the present work. We show that solutions of the stationary Dirac equation can be constructed for each internuclear distance $R$ by means of a two--step procedure. As will be discussed in Section~\ref{subsec:stationary_problem}, the use of the dual kinetically balanced (DKB) B--spline basis set method \cite{ShT04} for finding eigenfunctions of the monopole Hamiltonian constitutes the first step of the procedure. Based on the ``monopole'' basis set, which is, thanks to the DKB algorithm, free of spurious, non--physical solutions, we generate then, in the second step, the two--center wavefunctions for any required number of multipoles in the potential expansion. Since the effective solution of the stationary two--center problem is, by itself, only an intermediate stage in the treatment of the time--dependent Dirac equation, the evaluation of the wave packet describing the electron dynamics in the field of moving nuclei will be discussed in Section~\ref{subsect:time_dependent}. In particular, we obtain the decomposition of such a packet in terms of (stationary) two--center wavefunctions and determine the expansion coefficients. Although the developed approach can be applied to \textit{any} collision between bare and hydrogen--like ions, independent of their charges and impact parameter, here we restrict our analysis to two case studies of the electron loss in the course of (i) nuclear $\alpha$ decay and (ii) charge--symmetric ion--ion scattering at zero impact parameter. The first of these processes may be understood reasonably well within the framework of the first--order perturbation theory \cite{FiF77,Law77,AnA82,McA11} which will be employed in Sec.~\ref{subsec:alpha_decay} for testing the accuracy of our (non--perturbative) calculations. In contrast, the ionization accompanying slow collisions between two heavy ions provides an example of a purely non--perturbative problem. To demonstrate the potential of the proposed method for tackling this problem, we present in Sec.~\ref{subsec:results_UU} predictions for the $K$--shell ionization in the U$^{91+}$--U$^{92+}$ scattering. Based on the calculations conducted, we confirm a good performance of our time--dependent non--perturbative approach, provided that the full multipole expansion of the electron--nuclei interaction is taken into account. Summary of our results and a brief outlook will be given in Section~\ref{sec:summary}.

\medskip

Natural units ($\hbar$ = $m_{e}$ = c = 1) are used throughout the paper.

%
%

\section{Theoretical background}
\label{sec:theory}

The electron dynamics in the Coulomb field of two nuclei is described by the time--dependent Dirac equation:
\begin{equation}
   \label{eq:Dirac_equation}
   i\frac{\partial}{\partial t}\Psi(\bfr,t) = {\hat{H}}_{TC}\Psi(\bfr,t) \, ,
\end{equation}
where the Hamiltonian reads, in spherical coordinates, as:
\begin{equation}
   \label{eq:Dirac_Hamiltonian}
   {\hat{H}}_{TC} = \balpha\cdot\bfp + V(Z_1,|\bfr-\bfR_1|) + V(Z_2,|\bfr-\bfR_2|) + \beta \, .
\end{equation}
In this expression, $\bfp=-i\bnabla$ is the electron momentum operator, $\beta$ and ${\bm \alpha} = \{\alpha_x, \alpha_y, \alpha_z \}$ are the standard Dirac matrices, and the potential generated by the $i$th nucleus:
\begin{equation}
   \label{eq:potential}
   V(Z_i, |\bfr-\bfR_i|) = \alpha \int_{0}^{\infty} {\rm d}r' \frac{\rho(r',Z_i)}{\max(r,R_i)} \, ,
\end{equation}
is a function of its charge density distribution $\rho(r,Z_i)$ and charge $Z_i$. Moreover, $\bfR_1$ and $\bfR_2$ describe positions of the nuclei with respect to the center--of--mass of the system:
\begin{eqnarray}
   \label{eq:R1_R2}
   \bfR_1 &=& \frac{M_2}{M_1+M_2} \, \bfR \, , \nonumber \\
   \bfR_2 &=& - \frac{M_1}{M_1+M_2} \, \bfR ,
\end{eqnarray}
where the internuclear vector $\bfR \equiv \bfR(t)$ varies over time.

\medskip
In what follows, we shall discuss the solution of the time--dependent Dirac equation (\ref{eq:Dirac_equation})--(\ref{eq:potential}) for relative ion velocities that are much smaller than the bound electron velocity $v \approx \alpha Z_i$ . For such a slow collision regime, the adiabatic approach is justified and requires first the treatment of the \textit{static} two--center problem. In the next subsection, therefore, we will show how the eigensolutions of the time--independent (two--center) Hamiltonian can be efficiently generated for any internuclear distance.

\subsection{Stationary two--center Dirac problem}
\label{subsec:stationary_problem}

For each (instantaneous) position of the nuclei, the spectrum of the two--center system can be obtained by solving the time--independent Dirac equation:
\begin{equation}
   \label{eq:time-independent_problem}
   \hat{H}_{TC} \Phi(\bfr) = E \Phi(\bfr)
\end{equation}
where $E$ is the total energy and the Hamiltonian $\hat{H}_{TC}$ is given by Eq.~(\ref{eq:Dirac_Hamiltonian}). Analysis of such an eigenproblem, can be significantly simplified by the proper choice of the quantization ($z$--) axis. For example, by setting this axis along the internuclear vector ${\bm R}$, we can write the multipole expansion of the two--center potential from Eq.~(\ref{eq:Dirac_Hamiltonian}) in the form:
\begin{eqnarray}
   \label{eq:potential_expansion}
   V_{TC}(\bfr,\bfR) &=& V(Z_1,|\bfr-\bfR_1|) + V(Z_2,|\bfr-\bfR_2|) \nonumber \\
   &=& \sum\limits_{l=0}^\infty V_l(r,R) P_l(\cos \theta) \, ,
\end{eqnarray}
where $P_l$ is the Legendre polynomial, $\theta$ is the polar angle of the vector ${\bm r}$, and the expansion coefficients $V_l$ are given by:
\begin{eqnarray}
   \label{eq:expansion_coefficients}
    V_l(r,R) &=& \frac{2l+1}{2} \, \int\limits_{0}^\pi \sin \theta \, {\rm d}\theta \,
   \Big( V(Z_1,|\bfr-\bfR_1|) \nonumber \\
   &+& V(Z_2,|\bfr-\bfR_2|) \Big) P_l(\cos \theta)\,.
\end{eqnarray}
Moreover, if summation over $l$ in Eq.~(\ref{eq:expansion_coefficients}) is restricted to the zeroth term, $l =0$, the electron--nuclear interaction is governed by the spherically symmetric potential $V_{TC}(\bfr,\bfR) = V_0(r, R)$. The solution of the Dirac equation within such a \textit{monopole} approximation is well--elaborated and has been discussed in a number of works \cite{BeS76,ReM81,AcH07,AcH08,DeM12}. In particular, the eigenfunctions of the monopole Hamiltonian $\hat{H}_{TC}^{(0)} = \balpha\cdot\bfp + V_0(r, R) + \beta$ can be found in the form:
\begin{equation}
  \label{eq:wavefunction_monopole}
  \phi_{\kappa \mu}(\bfr)= \frac{1}{r} \left( \ba{c}
  G_\kappa(r)\chi_{\kappa \mu}({\hat{\bfr}}) \\ i
  F_\kappa(r)\chi_{-\kappa \mu}({\hat{\bfr}}) \end {array} \right) \, ,
\end{equation}
where $\chi_{\kappa \mu}$ is the standard Dirac spinor and the radial components satisfy the equation:
\begin{eqnarray}
   \label{eq:radial_system}
   \left[ \matrix{ V_0 + 1          & -\frac{{\rm d}}{{\rm d}r} + \frac{\kappa}{r}  \cr
           \frac{{\rm d}}{{\rm d}r} + \frac{\kappa}{r}    & V_0-1 } \right] \,
   \left( \matrix{G_\kappa(r) \cr F_\kappa(r)} \right) = \epsilon \left( \matrix{G_\kappa(r) \cr F_\kappa(r)} \right) \, .
\end{eqnarray}
In order to solve this radial eigenproblem, we use the dual kinetically balanced (DKB) B--spline basis set method \cite{ShT04}. Since such a DKB approach has been widely applied in the past for the treatment of spherically symmetric Dirac problems, we will not discuss its details here. Instead, we just mention that the DKB method allows one to avoid the spurious (non--physical) solutions of $\hat{H}_{TC}^{(0)}$ and to generate a quasi--complete set of wavefunctions $\{ \phi^n_{\kappa \mu}(\bfr) \}$, $n = 1, ... N$ for each value of the Dirac angular quantum number $\kappa$. These functions describe the electron states with the energy $\epsilon_{n \kappa}$ in the spherically--symmetric potential $V_0(r, R)$, and their overall number $N$ depends on the size of the basis set.

\medskip

The sets of eigenfunctions $\phi^n_{\kappa \mu}(\bfr)$, derived for the spherically symmetric problem can be employed to describe the electron dynamics for relatively small distances between colliding nuclei. If $R$ increases, the monopole approximation is no longer valid and one has to account for the full two--center potential (\ref{eq:potential_expansion}) when solving the time--independent Dirac equation (\ref{eq:time-independent_problem}). In the present work, we propose to present solutions of such an \textit{exact} eigenproblem in terms of the monopole functions:
\begin{equation}
   \label{eq:exact_wave_function_expansion}
   \Phi_{\mu}(\bfr) = \sum\limits_{n = 1}^N \sum\limits_{\kappa = -K}^{K} C_{n \mu}^\kappa \, \phi^n_{\kappa \mu}(\bfr) \, ,
\end{equation}
where $K$ is a parameter limiting the number of partial waves in the sum. The expansion coefficients $C_{n \mu}^\kappa$ can be determined then based on the principle of least action, $\delta {\mathcal S} = 0$, where the action is defined as:
\begin{equation}
   \label{eq:action}
   {\mathcal S} = \mem{\Phi_{\mu}}{\hat{H}_{TC} - E}{\Phi_{\mu}} \, .
\end{equation}
By inserting the wavefunction (\ref{eq:exact_wave_function_expansion}) into this expression and by evaluating the variation $\delta {\mathcal S}$ with respect to the change of expansion coefficients $C_{n \mu}^\kappa$, we obtain a system of differential equations:
\begin{equation}
   \label{eq:least_action_coefficients}
   \frac{\partial {\mathcal S}}{\partial C_{n \mu}^\kappa} = 0 \, ,
\end{equation}
which can be re--written in matrix form as follows:
\begin{equation}
   \label{eq:linear_algebra_problem}
   {\hat {\mathcal H}} \vec{C} = E \vec{C} \, .
\end{equation}
Upon evaluation of the elements of the matrix $\hat{\mathcal H}$:
\begin{equation}
   \label{eq:H_matrix_elements}
   {{\mathcal H}}_{i,k} = \epsilon_k \delta_{i,k} + \mem{\phi_i}{\sum\limits_{l=1}^{2K} V_l(r,R) P_l(\cos \theta)}{\phi_k} \, ,
\end{equation}
Eq.~(\ref{eq:linear_algebra_problem}) allows one to determine the vector $\vec{C} = \{ C_1, C_2, .... C_{N_{max}}\}$. Here, for the sake of brevity, we use short--hand notations $C_j \equiv C_{n \mu}^\kappa$, $\phi_j \equiv \phi^n_{\kappa \mu}$ and $\epsilon_j = \epsilon_{n \kappa}$.

\medskip

As seen from the discussion above, the spectrum of the time--independent Hamiltonian (\ref{eq:Dirac_Hamiltonian}) for each fixed internuclear distance $R$ can be generated by means of the two--step procedure. In the first step, we employ the DKB finite--basis set approach to find solutions
$\{ \epsilon_{n \kappa} \, , \phi^n_{\kappa \mu}\}$ of the monopole Hamiltonian. These solutions are used then, in the second step, to solve the generalized eigenvalue problem (\ref{eq:linear_algebra_problem}) and to obtain both the expansion coefficients $C_{n \mu}^\kappa$ of the wavefunctions $\Phi_{k \mu}(\bfr)$ and the energies $E_{k \mu}$ of the electron states in the \textit{full} two--center potential (\ref{eq:potential_expansion}). In the next Section, such a new set of eigenstates $\{ E_{k \mu}, \Phi_{k \mu} \}$ will be employed for solving the non--stationary Dirac problem.

\subsection{Time--dependent two--center Dirac problem}
\label{subsect:time_dependent}

Having generated a (quasi--) complete set of eigenstates of the two--center Hamiltonian (\ref{eq:Dirac_Hamiltonian}) at each internuclear distance $R$, we are ready now to solve the time--dependent equation (\ref{eq:Dirac_equation}) using the coupled channel method. Within this approach, the electron wavepacket $\Psi(\bfr,t)$ is expanded:
\begin{equation}
   \label{eq:Psi_expansion}
   \Psi(\bfr,t) = \sum\limits_{k \mu} a_{k \mu}(t) \Phi_{k \mu}(\bfr,t) \, ,
\end{equation}
in terms of the functions $\Phi_{k\mu}$ which parametrically depend on the internuclear distance and, hence, on time $t$. In Eq.~(\ref{eq:Psi_expansion}), moreover, $a_{k\mu}(t)$ are the time--dependent expansion coefficients whose squares, $|a_{k\mu}(t)|^2$, provide the occupation probabilities of the states $\ketm{\Phi_{k\mu}}$ at a particular instant in time. In order to find these coefficients, we substitute the expansion (\ref{eq:Psi_expansion}) into the Dirac equation (\ref{eq:Dirac_equation}) and derive the system of coupled channel equations:
\begin{eqnarray}
   \label{eq:CCE1}
   i \frac{{\rm d}}{{\rm d} t} a_{k \mu}(t) &=& E_{k \mu}(t)a_{k \mu}(t) \nonumber \\[0.1cm]
   && \hspace*{-1cm} - i\sum_{n\neq k, \, \mu'} a_{n \mu'}(t) \, \sprm{\Phi_{k \mu}(t)}{\frac{\partial \Phi_{n \mu'}(t)}{\partial t}} \, .
\end{eqnarray}
Any further analysis of this system requires the knowledge of how the electron--nuclei potential (\ref{eq:potential_expansion}) varies with time. Since the time--dependence enters into the problem solely through the internuclear distance $R$, the equation of motion of colliding nuclei must be established. In the present work we consider the simplest case of motion along the Rutherford trajectories. In this case the time, the internuclear distance and the tilt angle of the molecular axis can be expressed in terms of the dimensionless parameter $\xi$ as follows:
\begin{eqnarray}
   \label{eq:Rutherford_trajectory}
   t&=&\frac{a}{v_\infty}(\epsilon \sinh \xi + \xi) \,, \nonumber \\
   R&=&a(\epsilon \cosh \xi +1) \, , \nonumber \\
   \theta&=&2\arctan\left(\frac{\sqrt{\epsilon^2-1}
   \left(\tanh\left(\xi/2\right)+1\right)}{\epsilon+1-
   \left(\epsilon-1\right)\tanh\left(\xi/2\right)}\right) \, .
\end{eqnarray}
Here notations are introduced
\begin{eqnarray}
   a&=&\frac{\alpha Z_1 Z_2}{M_{12} v_\infty^2 }\, , \nonumber\\
   \epsilon&=&\left(1+\frac{b^2}{a^2}\right)^{1/2}\, ,
\end{eqnarray}
with $b$ denoting the impact parameter, $v_\infty$ the asymptotic value of the relative velocity of two particles at $t = \infty$, and $M_{12}$ the reduced mass. By inserting Eq.~(\ref{eq:Rutherford_trajectory}) into the system of coupled channel equations (\ref{eq:CCE1}) and re--writing it in terms of the parameter $\xi$, we derive:
\begin{eqnarray}
   \label{eq:CCE3}
   i \frac{{\rm d}}{{\rm d} \xi} a_{k \mu}(\xi) &=&
   \left(\frac{\partial t}{\partial \xi}\right) E_{k \mu}(\xi)a_{k \mu}(\xi) \nonumber \\[0.2cm]
   && \hspace*{-2cm} - i\sum_{n\neq k, \, \mu'} a_{n \mu'}(\xi)\left(
   \frac{\mem{\Phi_{k \mu}(\xi)}{\frac{\partial R}{\partial \xi} \frac{\partial }{\partial
   R}V(\bfr,\bfR,\xi)}{\Phi_{n \mu'}(\xi)}}{E_{n \mu'}(\xi)-E_{k \mu}(\xi)}\right. \nonumber \\
   [0.2cm]
   && \hspace*{-1cm} \left. -i\frac {d\theta}{d\xi}\mem{\Phi_{k \mu}(\xi)}{j_y}{\Phi_{n \mu'}(\xi)}\right) \, ,
\end{eqnarray}
where we used the relation
\begin{eqnarray}
   \label{eq:dphi_dt_replacement}
   \sprm{\Phi_{k \mu}}{\dot{\Phi}_{n \mu'}} &=&\bar{\delta}_{k,n}\frac{
   \mem{\Phi_{k \mu}}{\dot{R} \frac{\partial V_{TC}}{\partial R}}{\Phi_{n \mu'}}}{(E_{n \mu'} - E_{k \mu})}
   \nonumber \\ &&
   -i\frac{d \theta}{d t}\mem{\Phi_{k \mu}}{j_y}{\Phi_{n \mu'}}\,,
\end{eqnarray}
which is valid if the collision occurs in XZ--plane. Here $j_y$ is $y$--component of the total momentum projection operator and $\bar{\delta}_{k,n}$ is the anti--Kronecker delta symbol. The parametrization of $t$, $R$ and $\theta$ in terms of $\xi$ is most natural since the differential equation governing the time evolution of $R$ is autonomous i.e. an exact solution is possible only for $t(R)$ and not, as required, for $R(t)$.

\medskip

In order to solve the system of coupled channel equations and, hence, to find the expansion coefficients $a_k$, it is convenient to re--write Eq.~(\ref{eq:CCE3}) in matrix form:
\begin{equation}
   \label{eq:CCE_matrix_form}
    i\frac{\partial}{\partial \xi}\vec{a}(\xi) = M(\xi) \vec{a}(\xi),
\end{equation}
where $\vec{a} = \{a_1, a_2, ...\}$, and the individual elements of $M_{k,n}(\xi)$ are given by
\begin{eqnarray}
   \label{eq:M_matrix}
   M_{k \mu, \, n \mu'}(\xi)& =& \frac{\partial t}{\partial
   \xi} \, E_{k \mu} \, \delta_{k,n} \, \delta_{\mu \mu'} \nonumber \\
   &-& i\frac{\mem{\Phi_{k \mu}}{\frac{\partial R}{\partial
   \xi}\frac{\partial }{\partial
   R}{\hat{H}}_{TC}}{\Phi_{n \mu'}}}{E_{n \mu'} - E_{k \mu}}\bar{\delta}_{k,n} \, \delta_{\mu \mu'} \nonumber \\
  && -\frac{d \theta}{d \xi}\mem{\Phi_{k \mu}}{j_y}{\Phi_{n \mu'}}\, .
\end{eqnarray}
The matrix equation (\ref{eq:CCE_matrix_form}) can be integrated numerically on a grid of spacing $\Delta\xi$ according to:
\begin{equation}
   \label{eq:time_propagation}
   \vec{a}(\xi +\Delta \xi)= {\rm e}^{-i M(\xi+\frac{\Delta \xi}{2}) \cdot \Delta\xi} \, \vec{a}(\xi)
   + \mathcal{O}(\Delta\xi^3) \, ,
\end{equation}
and determines the vector $\vec{a}(\xi +\Delta \xi)$ at the ``time'' $\xi + \Delta\xi$ provided that the expansion coefficients coefficients $a_{k \mu}$ at the earlier moment $\xi$ are known. Since the matrix exponential in the right--hand--side of Eq.~(\ref{eq:time_propagation}) is unitary, the norm of the vector $\vec{a}$ will be preserved at each iteration.

\medskip
The iteration scheme (\ref{eq:time_propagation}) represents the final step in the numerical treatment of the time--dependent two--center Dirac equation (\ref{eq:Dirac_equation}). In Section \ref{sec:results}, we will use this scheme in order to investigate the electron ionization induced by the nuclear $\alpha$ decay as well as the slow ion--ion collisions. In the present calculations, we shall restrict ourselves to the simplest case of zero--impact--parameter picture, $b$ = 0. Within this framework, the last term in Eqs.~(\ref{eq:CCE3}) and (\ref{eq:dphi_dt_replacement}) vanishes and, hence, the matrix elements of the evolution matrix (\ref{eq:M_matrix}) are diagonal in $\mu$.

%
%
%
%
\section{Details of computations}
\label{sec:computations}

Having discussed the non--perturbative approach to the solution of the two--center Dirac problem, we are ready now to investigate the electron emission accompanying both, the $\alpha$--decay of heavy nuclei and the slow ion--ion collisions. Before starting with the presentation and analysis of the numerical results, let us briefly summarize the most important details of our calculations which, as mentioned in Section~\ref{sec:theory}, can be split into three stages. In the first step of this procedure, the eigenfunctions of the spherically--symmetric Hamiltonian $\hat{H}_{TC}^{(0)}$ are obtained by the DKB B--spline basis set method which guarantees the absence of the non--physical spurious states in the spectrum \cite{ShT04}. In the present work, we used about 200 B--splines of \textit{eighth} order defined in a box of size $L \simeq 10^5$ fm in order to construct ``monopole'' wavefunctions $\phi^n_{\kappa \mu}(\bfr)$ with energies in the range $0 \le \epsilon_{n \kappa} \le 10$ mc$^2$. Based on the detailed numerical analysis, we argue that such a truncated basis set allows one to achieve $\sim$~5--10~\% accuracy in the prediction of the ionization cross sections. As the second step of the non--perturbative treatment, the solutions of the \textit{full} two--center Hamiltonian (\ref{eq:Dirac_Hamiltonian}) are expanded in terms of $\phi^n_{\kappa \mu}(\bfr)$ (cf. Eq.~(\ref{eq:exact_wave_function_expansion})). Along this line, we obtain about 300 functions $\Phi_{k \mu}(\bfr; R)$ and corresponding energies $E_k$ for each internuclear distance R (or, equivalently, dimensionless parameter $\xi$ (\ref{eq:Rutherford_trajectory})). It is worth mentioning that the solutions of eigenproblem (\ref{eq:linear_algebra_problem}) and, hence, $\Phi_{k \mu}(\bfr)$ are defined up to an arbitrary sign. In our calculations, this sign is chosen for \text{all} $\Phi_{k \mu}(\bfr)$ from the requirement that their large radial components, calculated for two successive steps over $\xi$, behave similarly near the origin of the coordinates, i.e. for $r = 0 ... 500$~fm.

\medskip

With the help of generated basis sets $\{ \Phi_{k \mu}(\bfr; R(\xi)) \}_{k =1, ... N}$ we are finally able to perform the time propagation of the electron wavepacket in the field of moving nuclei. Prior to starting this propagation, one has to define the electron wavefunction in the initial moment of time. Indeed, the initial conditions depend on the particular process under consideration. For the nuclear $\alpha$--decay, for example, we assume that the electron is originally in the ground $1s_{1/2}$ state of the united nucleus of charge $Z$. Since the time--propagation begins from the moment when the $\alpha$ particle leaves the potential barrier at the distance $R_0 \sim$~10~fm from the daughter nucleus, we project, at $\xi = 0$ (corresponding to $t$ = 0), the wavefunction $\psi_{1s_{1/2}}(\bfr; Z)$ onto the basis set of eigenfunctions of Hamiltonian (\ref{eq:Dirac_Hamiltonian}) describing system of two Coulomb centers with charges $Z_1 = 2$ and $Z_2 = Z - 2$, placed at distances $R_1$ and $R_2$ with respect to their center--of--mass (see Eq.~(\ref{eq:R1_R2})). Such a projection procedure allows us to account for the shake--off effect and to obtain the first set of expansion parameters $\{a_{k \mu}(\xi = 0)\}_{k = 1, ... N}$ which are used then to find the electron wavepacket in subsequent time steps (cf. Eq.~(\ref{eq:time_propagation})). In order to produce results, presented in the Section \ref{subsec:alpha_decay}, time propagation was carried out for about 750 such steps of $\Delta\xi$ = 0.01, this corresponds to the retreat of the $\alpha$ particle to a distance of about 10$^4$~fm.

%
%
\begin{figure}[t]
\includegraphics[scale=0.4]{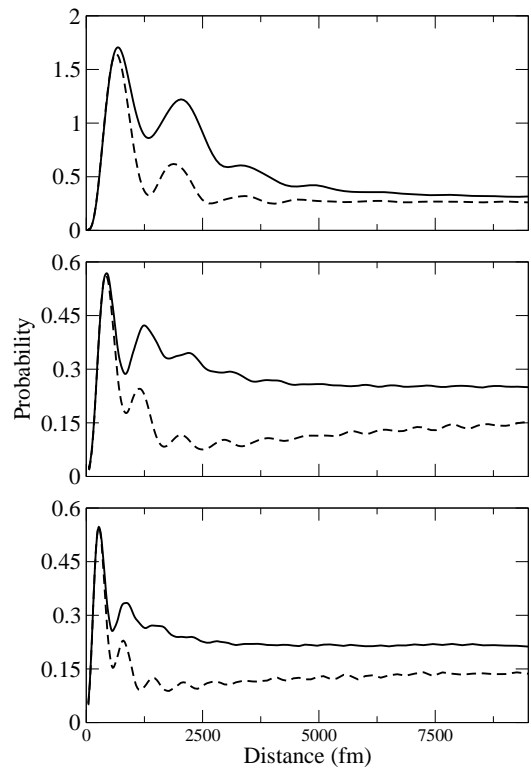}
\caption{$K$--shell ionization probability of hydrogen--like xenon (top panel), gadolinium (middle panel) and polonium (bottom panel) ions following the $\alpha$--decay. Non--perturbative calculations were carried out within the monopole approximation (dashed line) and by taking the full two--center potential into account (solid line). The probability is scaled $\times 10^5$.
\label{Fig1}}
\end{figure}

\medskip

In contrast to the $\alpha$--decay, the time propagation of the electron wavepacket in the field of two colliding uranium ions, studied in Section~\ref{subsec:results_UU}, was started from the moment when the ions are separated from each other by the distance $R = 5 \cdot 10^3$~fm. In this initial moment, the electron finds itself in the ground $1s_{1/2}$ state of one of the projectiles. The wavefunction of such a state is given by the sum of the lowest--lying \textit{gerade} and \textit{ungerade} solutions of the (stationary) two--center Hamiltonian, $\psi_{1s_{1/2}}(\bfr_i; Z = 92) \approx 1/\sqrt{2} \left( \Phi_{1\sigma_g} + \Phi_{1\sigma_u} \right)$; an approximation whose quality increases with the number of partial waves in the expansion (\ref{eq:exact_wave_function_expansion}). In the calculations bellow all partial waves with the Dirac angular quantum number in the range $\kappa = -10 ... +10$ are employed leading to about 10 \% accuracy of the presented results.

%
%
%
%
\section{Results and discussion}
\label{sec:results}

\subsection{Ionization following $\alpha$--decay of heavy nuclei}
\label{subsec:alpha_decay}

The non--perturbative approach presented in Section~\ref{sec:theory} can be used to study basic atomic processes accompanying slow collisions of two ions independent of their nuclear charges $Z_1$ and $Z_2$. In this section, we employ it to re--analyze the nuclear $\alpha$ decay, which is an example of a (charge--) asymmetric collision, $Z_1 << Z_2$, with \textit{zero} impact parameter and which can be treated also within first--order perturbation theory. As mentioned already, such a perturbative treatment has been successfully applied over the last decades in a large number of studies \cite{McA11,Law77,AnA82,ReN92}. In order to compare predictions of the non--perturbative and perturbative theories, we consider the decay of $\alpha$--active $^{110}$Xe, $^{148}$Gd and $^{210}$Po isotopes. For these zero--nuclear--spin nuclei, calculations have been performed for the ionization probability $P_{K}$ of an electron from the ground $1s_{1/2}$ state of an initially hydrogen--like system. In Fig.~\ref{Fig1} we display the non--perturbative results for the $P_K$ as a function of the internuclear distance. To deduce this probability, we have evaluated the electron wavefunction $\Psi(\bfr, t)$ at each step of the time propagation (see Eqs.~(\ref{eq:Psi_expansion})--(\ref{eq:time_propagation}) and related discussion) and projected it onto the positive--energy solutions of the two--center Dirac equation for the instantaneous distance $R = R(t)$
\begin{eqnarray}
   \label{eq:Pk_non-pert}
   P_K(R(t)) &=& \sum\limits_{E_k > mc^2} \left| \sprm{\Phi_{k \mu}(\bfr)}{\Psi(\bfr, t)} \right|^2  \nonumber \\
   &=& \sum\limits_{E_k > mc^2} \left| a_{k \mu}(t) \right|^2\, .
\end{eqnarray}
Calculations have been performed both within the monopole approximation, in which summation over $l$ in Eq.~(\ref{eq:potential_expansion}) is restricted to the zeroth term, and by taking the full two--center potential $V_{TC}(\bfr,\bfR)$ into account. As was expected, these two approaches agree only for relatively small internuclear distances. If  $R$ becomes greater than $500$~fm, the monopole approximation can significantly underestimate the ionization probability; an effect which becomes most pronounced for the heavy nuclei.

\medskip

Fig.~\ref{Fig1} shows that at very large distances, $R > 8000$~fm, the ionization probability $P_K$ converges to some final value which depends only on the charge of the mother nucleus and the initial velocity of the $\alpha$ particle. This ``asymptotic'' value of $P_K$ is displayed in Table~\ref{Tab1} for xenon, gadolinium and polonium ions, and compared with the results of our first--order perturbation calculations (see Ref.~\cite{McA11} for further details). Moreover, the previous (perturbative) predictions of Law \cite{Law77}, and Fischbeck and Freedman \cite{FiF75} obtained for the decay of polonium are given in the third column. As seen from the table, the non--perturbative treatment, based on the full multipole expansion of the two--center potential, reproduces well the ionization probabilities for all three ions. In particular, both perturbative and non--perturbative theories yield results that agree to within 5 \% if applied to the exploration of the $\alpha$ decay of polonium ions. If, however, the potential $V_{TC}(\bfr,\bfR)$ is approximated in Eq.~(\ref{eq:Dirac_Hamiltonian}) by the single monopole term, the non--perturbative calculations may result in approximately a 30 \% misestimation of $P_K$.

\begin{table}
\caption{\label{Tab1} $K$--shell ionization probability of hydrogen--like xenon, gadolinium and polonium ions following the $\alpha$--decay. The non--perturbative calculations, performed for $R \to \infty$ by using the monopole as well as exact approximations to the two--center potential, are compared with the first--order perturbation results and predictions by Law \cite{Law77}, and Fischbeck and Freedman \cite{FiF75}. The asymptotic kinetic energy of $\alpha$ particle $T_{\rm kin} = M_{\alpha} v^2_{\infty}/2$ from Ref.~\cite{Fir96} is given in the second column. All probabilities are of the order of $\times 10^6$.\\
}
\begin{tabular}{l....}
\hline
Ion \hspace{1.3cm}& \multicolumn{1}{c}{\hspace{0.3cm}$T_{\rm kin}$}\hspace{0.3cm}  &   \multicolumn{1}{c}{Perturbative}\hspace{0.3cm} & \multicolumn{2}{c}{Non--perturbative}  \\
\cline{4-5}
    &   \multicolumn{1}{c}{(MeV)}                    &              & \multicolumn{1}{c}{monopole}\hspace{0.2cm}& \multicolumn{1}{c}{exact} \\
\hline
$^{110}$Xe$^{+53}$	& 3.7 & 3.61	& 2.6 & 3.2\\[0.2cm]
$^{148}$Gd$^{+63}$	& 3.1 & 2.15	& 1.6 & 2.3\\[0.2cm]
$^{210}$Po$^{+83}$	& 5.4 & 2.00	& 1.4 & 2.1\\
                 &         & 1.81\footnotemark[1]   &      &      \\
                 &         & 2.03\footnotemark[2]   &      &      \\
\hline
\end{tabular}
\footnotetext[1]{Law \cite{Law77}} \footnotetext[2]{Fischbeck and Freedman \cite{FiF75}}
\end{table}

\medskip

\begin{figure}[t]
\includegraphics[scale=0.4]{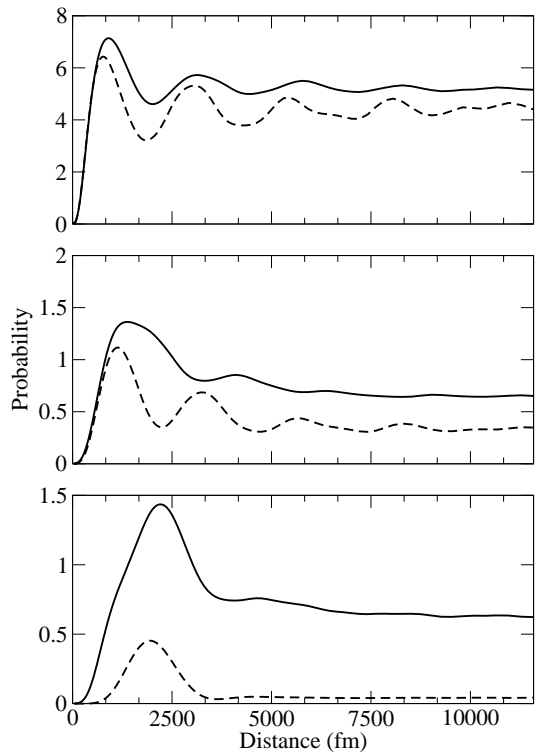}
\caption{Ionization probability of the $2s_{1/2}$ (top panel), $2p_{1/2}$ (middle panel), and $2p_{3/2}$ (bottom panel) states of hydrogen--like polonium $^{210}\mathrm{Po}^{+83}$ following the $\alpha$--decay. Non--perturbative calculations were carried out within the monopole approximation (dashed line) and by taking the full two--center potential into account (solid line). The probability is scaled $\times 10^5$.
\label{Fig2}}
\end{figure}

Until now we have discussed the $\alpha$--decay--induced ionization of hydrogen--like ions that have been \textit{prepared} initially in the ground $1s_{1/2}$ state. In order to verify the performance of the non--perturbative technique, based on the multipole expansion of the two--center interaction operator, it is also worth considering the electron emission from the various $L$ subshells. Even though experimental observation of the $L$--shell ionization of hydrogen--like systems might be hampered by the short lifetimes of excited ionic states, it can be measured for neutral atoms. Theoretically, such an atomic inner--shell ionization can be well described by using the developed approach if the proper screening potential is used in Eq.~(\ref{eq:potential}). The analysis of the screening effects in $\alpha$--decay--induced processes in neutral systems is, however, out of the scope of the present work. Instead, we just employ the $L$--shell ionization of hydrogen like ions as a testing ground for the non--perturbative theory from Section~\ref{sec:theory}. The internuclear--distance--dependent probabilities for the ionization of $2s_{1/2}$ (top panel), $2p_{1/2}$ (middle panel) and $2p_{3/2}$ (bottom panel) states of hydrogen--like polonium are evaluated based on this theory and are presented in Fig.~\ref{Fig2}. Similar to before, calculations have been performed by accounting for the full multipole expansion of the two--center potential (solid line) and by restricting this summation to the monopole term only (dashed line). Agreement between these two approaches can be observed again only for small internuclear distances, while for $R > 600 $~fm the monopole calculations underestimate the ionization probabilities by more than $25$\%. Moreover, the monopole approximation fails to reproduce $P_{2p_{3/2}}$ for the entire range of $R$.

\medskip

The asymptotic values of $P_{2s_{1/2}}$, $P_{2p_{1/2}}$ and $P_{2p_{3/2}}$ calculated for large distances $R$ are presented in Table~\ref{Tab2} and compared with the predictions of first--order perturbation theory \cite{McA11} and data by Law \cite{Law77}. As in the case of $K$--shell ionization, the full account of the electron--nuclei interaction $V_{TC}(\bfr,\bfR)$ in Eq.~(\ref{eq:Dirac_Hamiltonian}) leads here to approximately $5$\% agreement between the predictions of perturbative and non--perturbative theories for the entire $L$ shell. In contrast, the time propagation of the electron wavepacket in the spherically--symmetric potential $V_0(r, R)$ yields the probabilities $P_L$ that are $30$\% smaller comparing to the perturbative results. Again, these findings stress the importance of the higher multipole contributions to the electron--nuclei interaction for the time--dependent analysis (\ref{eq:time_propagation}) of the electron dynamics accompanying ion collisions.

%
%
%
%
%

%
%
%
\begin{table}
\caption{
\label{Tab2}
$L$--subshell ionization probabilities of hydrogen--like polonium $^{210}\mathrm{Po}^{+83}$ following the $\alpha$--decay. The non--perturbative calculations, performed for $R \to \infty$ by using the monopole as well as exact approximations to the two--center potential, are compared with the first--order perturbation results and predictions by Law \cite{Law77}. Probabilities given in units $\times 10^{5}$ and for the kinetic energy of the emerged $\alpha$--particle $T_{\rm kin}$ = 5.4~MeV \cite{Fir96}.}
\begin{tabular}{l...}
\hline
State     \hspace{1.4cm} &\multicolumn{1}{c}{\hspace{0.4cm}Perturbative}	\hspace{0.4cm}    &\multicolumn{2}{c}{Non--perturbative}  \\
\cline{3-4}
          &										 &\multicolumn{1}{c}{monopole}&\multicolumn{1}{c}{exact}\\
\hline
$2s_{1/2}$&               4.80                   &	4.0                            &   4.96 \\
          &               4.75\footnotemark[1]   &      					       &	  \\[0.2cm]
$2p_{1/2}$& 			  0.54				     & 0.30 					       &   0.64 \\
          & 			  0.50\footnotemark[1]   &  						       &      \\[0.2cm]
$2p_{3/2}$& 			  0.61				     & 0.04 					       &   0.61 \\
          & 			  0.60\footnotemark[1]   &  						       &      \\[0.2cm]
\hline
\end{tabular}
\footnotetext[1]{Law \cite{Law77}}
\end{table}
%
%
%
%

%
%
%
%
\subsection{Ionization in U$^{91+}$--U$^{92+}$ collisions}
\label{subsec:results_UU}

\begin{figure}[t]
\includegraphics[scale=0.3]{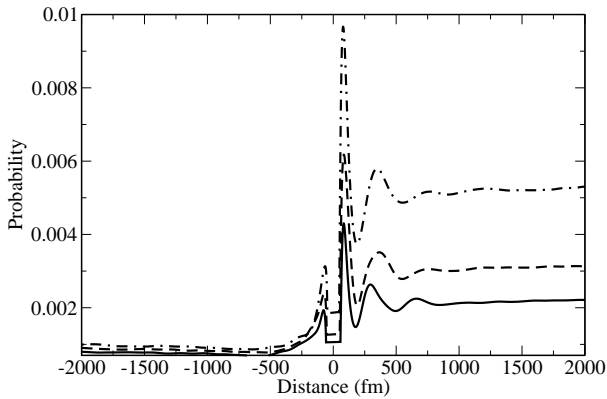}
\caption{Ionization probability of the $1s_{1/2}$ state of hydrogen--like uranium colliding with bare U$^{92+}$ ion. Non--perturbative calculations, based on the multipole expansion of the electron--nuclear interaction operator, were carried out for $zero$ impact parameter and for the initial relative kinetic energy of ions $T_p = $ 1.8 MeV/u (solid line), 2.0 MeV/u (dashed line), and 2.2 MeV/u (dash--dotted line). The negative and positive values of $R$ correspond to the times when ions approach and move away from each other, respectively.
\label{Fig3}}
\end{figure}

So far, we have shown that the time--dependent method (\ref{eq:CCE3}), based on the expansion of the basis wavefunctions in terms of monopole solutions, can be successfully utilized to study the $\alpha$--decay--induced ionization. Besides this---purely perturbative---problem, the performance of the developed approach has been also examined for slow collisions between two high--$Z$ ions. In contrast to the $\alpha$--decay, theoretical analysis of such collisions usually can not be carried out within the framework of the perturbation theory and \textit{demands} the application of non--perturbative techniques. Along this line we have focused, in particular, on the $K$--shell ionization in U$^{91+}$--U$^{92+}$ collisions at $zero$ impact parameter. The ionization probability $P_K$ has been calculated based on Eq.~(\ref{eq:Pk_non-pert}), where the electron wavepacket $\Psi(\bfr, t)$ was propagated from a time when the ions were at a distance $R = 5 \cdot 10^3$~fm, through the closest approach $R_0 \approx 50$~fm, to a moment when the internuclear distance increased again to $R = 5 \cdot 10^3$~fm. In Fig.~\ref{Fig3}, for example, $P_K$ is displayed as a function of the distance $R$ and for the (relative) collision energies $T_p$ = 1.8, 2.0 and 2.2 MeV/u. As seen from the figure, the steep rise of the ionization probability appears immediately after the point of closest approach $R_0$ at which the (relative) ionic motion is suddenly reversed and the electron can be ``shaken off'' into the continuum. Such a behaviour of the $P_K$ as well as its further damped oscillations have been predicted previously in Ref.~\cite{BeS76} based on the monopole approximation and now is confirmed by our theory that accounts for the multipole expansion of the electron--nuclei interaction. Moreover, our calculations clearly indicate a rise of the ionization probability with the collision energy. For example, the asymptotic value of the $P_K$ is increased by almost factor of \textit{three} if the initial (relative) energy changes from 1.8 to 2.2 MeV/u. Further significant enhancement of the $P_K$ is predicted for higher energies at which the ``diving'' of the ground quasi--molecular state into the Dirac's negative continuum takes place \cite{BeS76}. However, since the analysis of such strong--field phenomena is out of scope of the present paper, we restrict here our calculations to the ``undercritical'' energy range, $T_p \lesssim 2.3$ MeV/u.

%
%
%
%
\section{Summary and outlook}
\label{sec:summary}

In summary, we have laid out a theoretical approach to the time--dependent two--center Dirac problem. Within such an approach, the wavefunctions, describing the (single) electron dynamics in the field of two moving nuclei, are expanded in terms of solutions of the stationary Dirac equation (\ref{eq:time-independent_problem}). We have argued that these stationary solutions can be efficiently constructed in \textit{spherical} coordinates and for each internuclear distance $R$ by means of the two--step procedure. The first step of the procedure consists of employing the dual kinetically balanced (DKB) B--spline basis set method to find eigenfunctions of the Hamiltonian $\hat{H}_0$, which accounts for the spherically--symmetric part of the electron--nuclei interaction. On the basis of these functions we generate, in the second step, the required solutions of the stationary two--center problem.

%
%

\medskip

The developed time--dependent approach can help to explore various atomic processes accompanying slow ion collisions. In the present work, for example, we used this theory to calculate the electron--loss probabilities for the (i) $\alpha$--decay of hydrogen--like xenon, gadolinium and polonium ions, and (ii) U$^{91+}$--U$^{92+}$ scattering at zero impact parameter. The $\alpha$--decay, being an example of charge--asymmetric collisions, can be described sufficiently well within the framework of first--order perturbation theory. Calculations based on this theory have been used to prove the accuracy of our non--perturbative approach. For the $K$-- and $L$--shell $\alpha$--decay--induced ionization, predictions of both perturbative and non--perturbative methods were found to agree to within about 5 \% if the multipole expansion of the two--center potential is taken into account in the time--dependent Hamiltonian (\ref{eq:Dirac_Hamiltonian}). If, in contrast, this potential is approximated by its monopole term, our calculations may underestimate the ionization probabilities by more than 30 \%; this failure of the monopole approximation becomes most pronounced for large internuclear distances. Based on these findings we stressed the vital importance of the proper treatment of the electron--nuclei interaction for the accurate description of slow ion--ion collisions. The rigorous ``multipole'' approach has been employed then to explore the $K$--shell ionization accompanying U$^{91+}$--U$^{92+}$ collisions.  For this---purely non--perturbative---process, we qualitatively confirmed the impact--parameter--behaviour of the ionization probability, which was predicted previously by Betz and co-authors \cite{BeS76} within the monopole theory.

\medskip

Both the $\alpha$--decay of hydrogen--like heavy ions and the U$^{91+}$--U$^{92+}$ scattering have been explored in the present work for the case of \textit{zero} impact parameter. Of course, the developed non--perturbative method is not limited to such a simple geometry and can be applied to analyze heavy--ion collisions at $b \ne 0$. For these collisions, the last term of Eqs.~(\ref{eq:CCE3})--(\ref{eq:dphi_dt_replacement}), that accounts for the rotation of the internuclear distance, does not vanish and makes the elements of the evolution matrix (\ref{eq:M_matrix}) non-diagonal in $\mu$. The impact--parameter dependence of the electron loss as well as the excitation and the charge transfer processes will be discussed in a forthcoming publication and will help in planning future experiments on slow collisions between two high--$Z$ projectiles. These experiments are likely to be carried out at the Facility for Antiproton and Ion Research (FAIR) in Darmstadt and are expected to reveal unique information about the quantum electrodynamics of extremely strong fields.

%
%
%
%
\section{Acknowledgments}

S.M. acknowledges the support of the International Max Planck Research School for Quantum Dynamics (IMPRS--QD). This work is supported by the Helmholtz Gemeinschaft (Nachwuchsgruppe VH--NG--421).

%
%
%
%

\end{document}